\documentclass[journal]{IEEEtran}%
\pdfoutput=1
\usepackage{graphics,amsmath,amsfonts,amscd,revsymb,latexsym,
enumerate,multirow,epsfig}
\usepackage{amsmath}
\usepackage{amsfonts}
\usepackage{amssymb}
\usepackage{graphicx}%
\providecommand{\U}[1]{\protect\rule{.1in}{.1in}}
\providecommand{\U}[1]{\protect\rule{.1in}{.1in}}
\newtheorem{theorem}{Theorem}

\newcommand{\qed}{{\hfill$\Box$}}
\newenvironment{proof}{\noindent \textbf{{Proof~} }}{\qed}

\begin{document}

\title{Entanglement generation with a quantum channel and a
shared state}
\author{Mark M. Wilde and Min-Hsiu Hsieh\thanks{Mark M. Wilde was with the Electronic
Systems Division, Science Applications International Corporation, 4001 North
Fairfax Drive, Arlington, Virginia, USA\ 22203 when conducting this research,
but is now a postdoctoral fellow with the School of Computer Science at
McGill University. Min-Hsiu Hsieh is with the
ERATO-SORST Quantum Computation and Information Project, Japan Science and
Technology Agency 5-28-3, Hongo, Bunkyo-ku, Tokyo, Japan(E-mail:
mark.wilde@mcgill.ca and minhsiuh@gmail.com)}}
\maketitle

\begin{abstract}
We introduce a new protocol, the channel-state coding protocol, to quantum
Shannon theory. This protocol generates entanglement between a sender and
receiver by coding for a noisy quantum channel with the aid of a noisy shared
state. The mother and father protocols arise as special cases of
the channel-state coding protocol, where the channel is noiseless or the state
is a noiseless maximally entangled state, respectively. The channel-state
coding protocol paves the way for formulating entanglement-assisted quantum
error-correcting codes that are robust to noise in shared entanglement.
Finally, the channel-state coding protocol leads to a Smith-Yard
superactivation, where we can generate entanglement using a zero-capacity
erasure channel and a non-distillable bound entangled state.

\end{abstract}


\section{Introduction}

Quantum Shannon theory is the study of the ultimate capability of a noisy
quantum system to preserve correlations \cite{DHW05RI,Yard05a}. A noisy
quantum channel possesses various capacities: its quantum capacity governs its
ability to transmit quantum information \cite{Lloyd96,Shor02,Devetak03}, its
classical capacity governs its ability for noiseless classical communication
\cite{Hol98,SW97}, and its private capacity governs its ability for noiseless
private communication \cite{Devetak03,CWY04}. A noisy bipartite state
possesses various distillation yields. Its distillable entanglement determines
the amount of maximal entanglement that we can recover from it
\cite{BDSW96,DW03c,DW03b}. Its distillable secret key determines
its private correlations \cite{DW03c,DW03b}, and its distillable common
randomness determines its classical correlations \cite{DW03a}.

In their pioneering unification of quantum Shannon theory \cite{DHW03,DHW05RI}%
, Devetak \textit{et al}. formulated the mother and father protocols. The
mother protocol exploits a noisy bipartite state and noiseless quantum
communication to establish noiseless entanglement between two parties. The
father protocol exploits a noisy quantum channel and noiseless entanglement to
transmit noiseless quantum information from a sender to a receiver. Since this
work, various authors have unified quantum Shannon theory in other ways
\cite{ADHW06FQSW,arx2008oppenheim,HW09T3,WH10}.

In this paper, we introduce a new protocol, the \textit{channel-state coding
protocol}, that exploits both a noisy bipartite state and a noisy quantum
channel to establish noiseless entanglement between two parties. In the
operation of the independent and identically distributed (IID) version of the
protocol, we assume that the sender and receiver use the channel and the state
\textit{the same number of times}. One might think that the optimal strategy
is one of the following three strategies:

\begin{enumerate}
\item Distill entanglement from the state and generate entanglement with a
quantum channel code. The total entanglement generated is then the sum of the
distilled entanglement and the entanglement generated from the channel.

\item Distill entanglement and perform the father protocol if enough
entanglement is available. The amount of entanglement generated is the net
amount that the father protocol can generate.

\item Perform quantum channel coding and follow with the mother protocol if
enough quantum communication is available. The amount of entanglement
generated is the net amount that the mother protocol can generate.
\end{enumerate}
It turns out that none of the above strategies is the best strategy. The
channel-state coding protocol is the best strategy here and instead encodes
both the input to the quantum channel and a share of the noisy bipartite state.

The existence of the channel-state coding protocol has some interesting
ramifications. First, it addresses a practical concern for the theory of
entanglement-assisted coding \cite{BDH06}, where a sender exploits noiseless
entanglement and a noisy quantum channel to transmit quantum information. It
is conjectured, but not yet demonstrated, that the performance of an
entanglement-assisted code decreases dramatically if the entanglement is not
perfect. The channel-state coding protocol demonstrates that another strategy
other than entanglement-assisted coding is appropriate for this situation. In
fact, the motivation for the channel-state coding protocol was the discovery
of an entanglement-assisted code that corrects errors on both the sender's
transmitted qubits and the receiver's share of the entanglement
\cite{prep2007shaw}. Secondly, the mother and father protocols now arise as a
special case of the channel-state coding protocol. The mother arises when the
quantum channel that connects sender to receiver is a perfect quantum channel.
The father arises when the shared entanglement between sender and receiver is
perfect. Finally, it leads to another instance of the superactivation effect
\cite{science2008smith,SST03,LWZG09}. Specifically, we can apply the
Smith-Yard superactivation \cite{science2008smith}\ to show that it is
possible to establish entanglement using a quantum channel with zero capacity
and a noisy bipartite state with zero distillable entanglement.

We structure this paper as follows. We first outline a general channel-state
coding protocol. Section~\ref{sec:capacity-theorem}\ gives the proof of the
channel-state coding capacity theorem. This theorem determines the ultimate
rate at which a noisy quantum channel and a noisy state can generate maximal
entanglement. We then show how a special case of this protocol, doing quantum
channel coding and entanglement distillation, is inferior to
the channel-state coding protocol.
Section~\ref{sec:mother-father} shows how the father and mother protocol are
special cases of the channel-state coding protocol. We then show how it is
possible to obtain a Smith-Yard-like superactivation in the channel-state
coding protocol and conclude with some observations and open questions.

\section{Channel-State Coding Protocol}

We begin by defining our channel-state coding protocol for a quantum channel
$\mathcal{N}^{A_{1}^{\prime}\rightarrow B_{1}}$ and a noisy bipartite state
$\rho^{A_{2}B_{2}}$. The noisy quantum channel $\mathcal{N}^{A_{1}^{\prime
}\rightarrow B_{1}}$ connects a sender Alice to a receiver Bob, and Alice and
Bob share the noisy state $\rho^{A_{2}B_{2}}$ before the protocol begins. The
channel has an extension to an isometry $U_{\mathcal{N}}^{A_{1}^{\prime
}\rightarrow B_{1}E_{1}}$, defined on a bipartite quantum system $B_{1}E_{1}$.
Bob has access to system $B_{1}$ and Eve has access to system $E_{1}$. The
noisy state admits a purification $\psi^{A_{2}B_{2}E_{2}}$ where Eve shares
a purifying system $E_{2}$.

We appeal to the asymptotic setting where Alice and Bob have access to $n$
independent uses of the channel $\mathcal{N}^{A_{1}^{\prime}\rightarrow B_{1}%
}$ and $n$ copies of the bipartite state $\rho^{A_{2}B_{2}}$ (where $n$ is as
large as we need it to be). We denote the $n$ independent uses of the channel
as%
\[
\mathcal{N}^{A_{1}^{\prime n}\rightarrow B_{1}^{n}}\equiv(\mathcal{N}%
^{A_{1}^{\prime}\rightarrow B_{1}})^{\otimes n},
\]
and the $n$ copies of the bipartite state $\rho^{A_{2}B_{2}}$ as%
\[
\rho^{A_{2}^{n}B_{2}^{n}}\equiv(\rho^{A_{2}B_{2}})^{\otimes n}.
\]
Let $U_{\mathcal{N}}^{A_{1}^{\prime n}\rightarrow B_{1}^{n}E_{1}^{n}}$ and
$\psi^{A_{2}^{n}B_{2}^{n}E_{2}^{n}}$ similarly denote the $n^{\text{th}}$
extensions of the respective isometry $U_{\mathcal{N}}^{A_{1}^{\prime
}\rightarrow B_{1}E_{1}}$ and purification $\psi^{A_{2}B_{2}E_{2}}$.

Alice's task is to generate noiseless entanglement between her and Bob by
using the channel $\mathcal{N}^{A_{1}^{\prime n}\rightarrow B_{1}^{n}}$ and
the noisy state $\rho^{A_{2}^{n}B_{2}^{n}}$. At the end of the protocol, the
generated entanglement should be close to the following state:%
\begin{equation}
\left\vert \Phi\right\rangle ^{AB}\equiv\frac{1}{\sqrt{D}}\sum_{i=1}%
^{D}\left\vert i\right\rangle ^{A}\left\vert i\right\rangle ^{B},
\label{eq:desired-state}%
\end{equation}
where $D$ indicates the amount of entanglement generated so that the rate of
entanglement generation is $R\equiv\log\left(  D\right)  /n$. We allow the
free use of a forward classical channel from Alice to Bob.

An $(n,R,\epsilon)$ \textit{channel-state code} consists of three steps:
preparation, transmission, and channel decoding. We detail each of these steps below.

\textbf{Preparation.} Alice possesses her share of the noisy bipartite state
$\rho^{A_{2}^{n}B_{2}^{n}}$ with purification $\psi^{A_{2}^{n}B_{2}^{n}%
E_{2}^{n}}$. Alice employs a preparation map $\mathcal{P}^{A_{2}%
^{n}\rightarrow A_{1}^{n}A_{2}^{n}A_{1}^{\prime n}}$ that prepares the
following state for input to the channel:%
\[
\omega^{A_{1}^{n}A_{2}^{n}A_{1}^{\prime n}B_{2}^{n}E_{2}^{n}}\equiv
\mathcal{P}^{A_{2}^{n}\rightarrow A_{1}^{n}A_{2}^{n}A_{1}^{\prime n}}%
(\psi^{A_{2}^{n}B_{2}^{n}E_{2}^{n}}),
\]
where $A_{1}^{n}$ and $A_{2}^{n}$ are systems with the same respective
dimension as the input to the channel and Alice's share of the state.%

\begin{figure}
[ptb]
\begin{center}
\includegraphics[
natheight=2.866000in,
natwidth=7.360400in,
height=1.3725in,
width=3.0831in
]%
{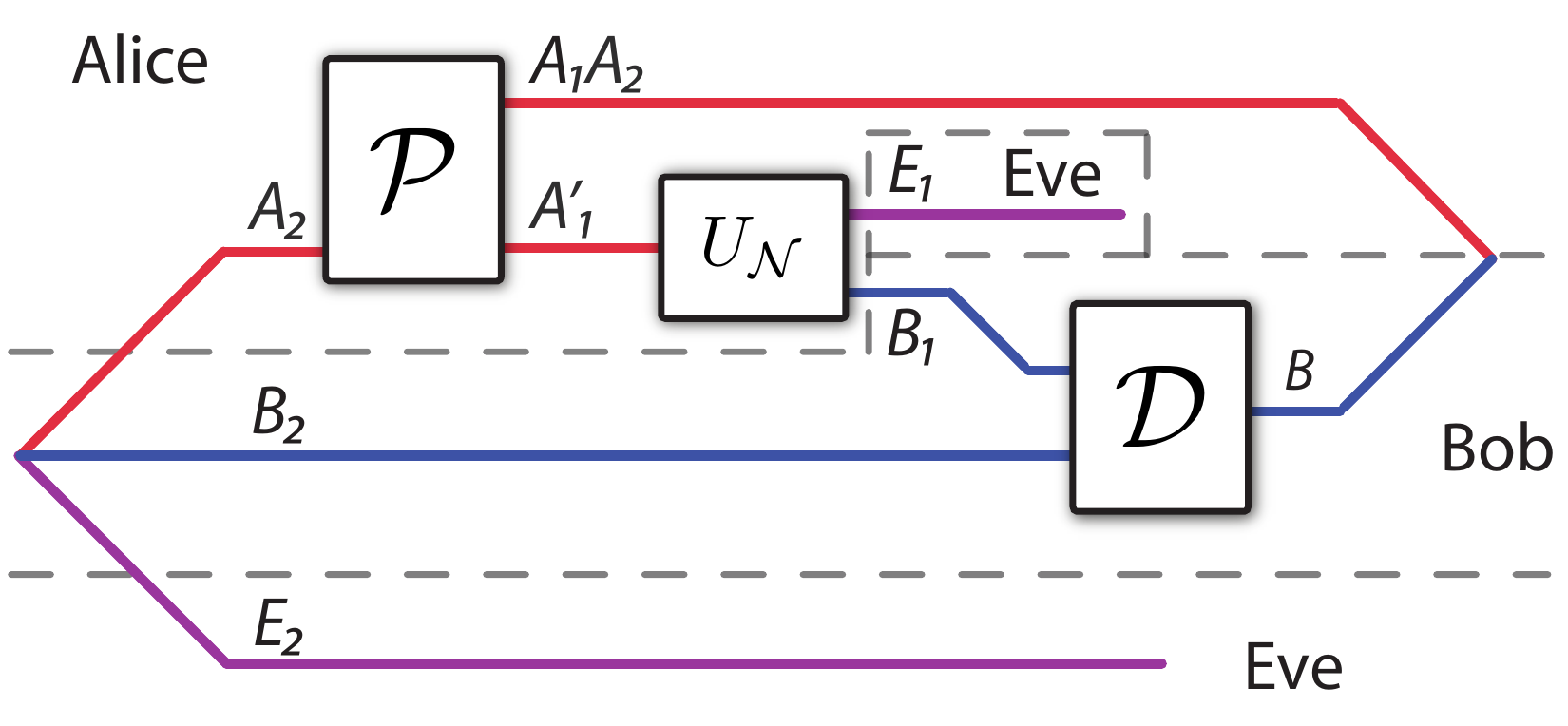}%
\caption{(Color online)\ The most general protocol for generating entanglement
with a noisy state $\rho^{A_{2}B_{2}}$ and a noisy channel $\mathcal{N}%
^{A_{1}^{\prime}\rightarrow B_{1}}$. It is implicit in the above diagrams that
the systems $A_{1}$, $A_{1}^{\prime}$, $B_{1}$, $E_{1}$, $A_{2}$, $B_{2}$,
$E_{2}$ are actually the respective $n$-copy systems $A_{1}^{n}$,
$A_{1}^{\prime n}$, $B_{1}^{n}$, $E_{1}^{n}$, $A_{2}^{n}$, $B_{2}^{n}$,
$E_{2}^{n}$. A good protocol generates the maximally-entangled state
$\Phi^{AB}$ shared between Alice and Bob.}%
\label{fig:converse-noisy-father}%
\end{center}
\end{figure}
\textbf{Transmission.} Alice sends the $A_{1}^{\prime n}$ system of the state
$\omega^{A_{1}^{n}A_{2}^{n}A_{1}^{\prime n}B_{2}^{n}E_{2}^{n}}$ through the
channel $U_{\mathcal{N}}^{A_{1}^{\prime n}\rightarrow B_{1}^{n}E_{1}^{n}}$.
This transmission generates the state%
\begin{equation}
\omega^{A_{1}^{n}A_{2}^{n}B_{1}^{n}E_{1}^{n}B_{2}^{n}E_{2}^{n}}\equiv
U_{\mathcal{N}}^{A_{1}^{\prime n}\rightarrow B_{1}^{n}E_{1}^{n}}(\omega
^{A_{1}^{n}A_{2}^{n}A_{1}^{\prime n}B_{2}^{n}E_{2}^{n}}%
).\label{eq:transmitted-state}%
\end{equation}

\textbf{Channel Decoding.} Bob receives the above state from the channel and
performs a decoding map $\mathcal{D}^{B_{1}^{n}B_{2}^{n}\rightarrow B}$
resulting in the state%
\begin{equation}
\left(  \omega^{\prime}\right)  ^{A_{1}^{n}A_{2}^{n}BE_{1}^{n}E_{2}^{n}}%
\equiv\mathcal{D}^{B_{1}^{n}B_{2}^{n}\rightarrow B}(\omega^{A_{1}^{n}A_{2}%
^{n}B_{1}^{n}E_{1}^{n}B_{2}^{n}E_{2}^{n}}). \label{eq:decoded-state}%
\end{equation}
The ideal state after Bob's decoding map is close in trace distance to the
following product state:%
\[
\Phi^{AB}\otimes\sigma^{E_{1}^{n}E_{2}^{n}},
\]
where $\Phi^{AB}$ is the state in\ (\ref{eq:desired-state}), the subspaces of
the systems $A_{1}^{n}A_{2}^{n}$ in which the entanglement is encoded are
isomorphic to system $A$ (Alice can perform some isometry to map between these
spaces), and $\sigma^{E_{1}^{n}E_{2}^{n}}$ is some constant state on Eve's
systems $E_{1}^{n}E_{2}^{n}$. The criterion for a successful channel-state
code is that%
\[
\left\Vert \left(  \omega^{\prime}\right)  ^{A_{1}^{n}A_{2}^{n}BE_{1}^{n}%
E_{2}^{n}}-\Phi^{AB}\otimes\sigma^{E_{1}^{n}E_{2}^{n}}\right\Vert _{1}%
\leq\epsilon,
\]
where $\epsilon > 0$. Figure~\ref{fig:converse-noisy-father}\ depicts all of the above steps in a
general channel-state coding protocol.

\section{The Channel-State Capacity Theorem}

\label{sec:capacity-theorem}A rate $R$ is \textit{achievable} if there exists
an $(n,R-\delta,\epsilon)$ channel-state code for any $\epsilon,\delta>0$ and
sufficiently large $n$.

\begin{theorem}
\label{thm:NFP}The entanglement generation capacity $E(\mathcal{N}\otimes
\rho)$ of a quantum channel $\mathcal{N}$ and a bipartite state $\rho$ is%
\begin{equation}
E(\mathcal{N}\otimes\rho)=\lim_{l\rightarrow\infty}\frac{1}{l}E^{(1)}%
(\mathcal{N}^{\otimes l}\otimes\rho^{\otimes l}), \label{eq:upper-bound}%
\end{equation}
where the \textquotedblleft one-shot\textquotedblright\ capacity
$E^{(1)}(\mathcal{N}\otimes\rho)$ is%
\begin{equation}
E^{(1)}(\mathcal{N}\otimes\rho)=\max_{\mathcal{P}}I\left(  A_{1}A_{2}\rangle
B_{1}B_{2}\right)  _{\omega}. \label{eq:one-shot-cap}%
\end{equation}
The maximization is over all preparations $\mathcal{P}^{A_{2}\rightarrow
A_{1}A_{2}A_{1}^{\prime}}$ and the coherent information $I\left(  A_{1}%
A_{2}\rangle B_{1}B_{2}\right)  _{\omega}$ is with respect to the following
state:%
\begin{equation}
\mathcal{N}^{A_{1}^{\prime}\rightarrow B_{1}}(\mathcal{P}^{A_{2}\rightarrow
A_{1}A_{2}A_{1}^{\prime}}(\rho^{A_{2}B_{2}})). \label{eq:maximization-state}%
\end{equation}

\end{theorem}

The proof of the above capacity theorem consists of two parts. The first part
that we prove is the \textit{converse theorem}. The multi-letter converse theorem states
that the capacity in the above theorem is optimal---any given coding scheme
that has asymptotically good performance cannot perform any better than the
rate in (\ref{eq:upper-bound}). The second part that we prove is the
\textit{direct coding theorem}. The proof of the direct coding theorem gives a
coding scheme that achieves the capacity in (\ref{eq:upper-bound}).

\begin{proof}
[Converse]We provide an upper bound on the entanglement generation rate $R$ of
a general channel-state coding protocol that allows the help of
classical communication. Consider the following chain of
inequalities:%
\begin{align*}
nR  &  =I\left(  A\rangle B\right)  _{\Phi}\\
&  =I\left(  A_{1}^{n}A_{2}^{n}\rangle B\right)  _{\Phi}\\
&  \leq I\left(  A_{1}^{n}A_{2}^{n}\rangle B\right)  _{\omega^{\prime}%
}+\epsilon^{\prime}\\
&  \leq I\left(  A_{1}^{n}A_{2}^{n}\rangle B_{1}^{n}B_{2}^{n}\right)
_{\omega}+\epsilon^{\prime}.
\end{align*}
The first and second equalities result from evaluating the coherent
information of the state $\Phi^{AB}$ and realizing that the system $A$ and the
subspace of $A_{1}^{n}A_{2}^{n}$ where Alice encodes the entanglement are
isomorphic. The maximally entangled state $\Phi^{AB}$ and $\left(
\omega^{\prime}\right)  ^{A_{1}^{n}A_{2}^{n}B}$ in (\ref{eq:decoded-state})
are $\epsilon$-close for a good code. Noting this fact, the first inequality
results from an application of the Alicki-Fannes' inequality
\cite{0305-4470-37-5-L01} where $\epsilon^{\prime}\equiv4\epsilon\log
D+2H(\epsilon)$, $H(\epsilon)$ is the binary entropy function, and
$\lim_{\epsilon\rightarrow0}H(\epsilon)=0$. The last inequality results from
quantum data processing \cite{SN96}, where we evaluate the coherent
information with respect to the state $\omega^{A_{1}^{n}A_{2}^{n}B_{1}%
^{n}E_{1}^{n}B_{2}^{n}E_{2}^{n}}$ in (\ref{eq:transmitted-state}). The
converse theorem holds because the state $\omega^{A_{1}^{n}A_{2}^{n}B_{1}%
^{n}E_{1}^{n}B_{2}^{n}E_{2}^{n}}$ is a state of the form
(\ref{eq:maximization-state}).
\end{proof}

We can phrase the direct coding theorem as a resource inequality
\cite{DHW05RI}:%
\begin{equation}
\left\langle \mathcal{N}\otimes\rho\right\rangle +I\left(  A_{1}A_{2}%
;E_{1}E_{2}\right)  _{\omega}\left[  c\rightarrow c\right]  \geq I\left(
A_{1}A_{2}\rangle B_{1}B_{2}\right)  _{\omega}\left[  qq\right]
.\label{eq:resource-inequality}%
\end{equation}
The above resource inequality is an asymptotic statement of achievability.
Suppose Alice has access to $n$ independent uses of the noisy quantum channel
$\mathcal{N}$, $n$ shares of $n$ respective copies of the noisy bipartite
state $\rho$, and $nI\left(  A_{1}A_{2};E_{1}E_{2}\right)  _{\omega}$ bits of
classical communication. Then she can reliably generate $nI\left(  A_{1}%
A_{2}\rangle B_{1}B_{2}\right)  _{\omega}$ ebits of entanglement with Bob. The
entropic quantities are with respect to the state%
\[
\mathcal{N}^{A_{1}^{\prime}\rightarrow B_{1}}(\mathcal{P}^{A_{2}\rightarrow
A_{1}A_{2}A_{1}^{\prime}}(\rho^{A_{2}B_{2}})),
\]
where $\mathcal{P}^{A_{2}\rightarrow A_{1}A_{2}A_{1}^{\prime}}$ is a
preparation operation equivalent to appending the state $\rho^{A_{2}B_{2}}$
with a state $\phi^{A_{1}A_{1}^{\prime}}$ and performing an isometric encoding
$\mathcal{E}^{A_{1}^{\prime}A_{2}\rightarrow A_{1}^{\prime}A_{2}}$ so that%
\[
\mathcal{P}^{A_{2}\rightarrow A_{1}A_{2}A_{1}^{\prime}}(\rho^{A_{2}B_{2}%
})=\mathcal{E}^{A_{1}^{\prime}A_{2}\rightarrow A_{1}^{\prime}A_{2}}%
(\phi^{A_{1}A_{1}^{\prime}}\otimes\rho^{A_{2}B_{2}}).
\]
We are specifically counting the classical communication cost in the above
resource inequality and show how the amount in (\ref{eq:resource-inequality})
arises in the proof of the theorem.

The proof of the direct coding theorem exploits the observations and coding
techniques from Refs.~\cite{qcap2008first,ADHW06FQSW}. We refer the reader to
these papers for details of carrying out the proof.

We first establish some notation and concepts for the proof. Alice has many
uses of the channel $\mathcal{N}^{A_{1}^{\prime}\rightarrow B_{1}}$ available,
and we label this channel as $\mathcal{N}_{1}^{A_{1}^{\prime}\rightarrow
B_{1}}$ (with a subscript \textquotedblleft1\textquotedblright) for reasons
that become clear later. Let $U_{\mathcal{N}_{1}}^{A_{1}^{\prime}\rightarrow
B_{1}E_{1}}$ denote an isometric extension of the channel $\mathcal{N}%
_{1}^{A_{1}^{\prime}\rightarrow B_{1}}$. Suppose Alice prepares the state
$\phi^{A_{1}A_{1}^{\prime}}$ on the systems $A_{1}A_{1}^{\prime}$. Sending the
$A_{1}^{\prime}$ system of the state $\phi^{A_{1}A_{1}^{\prime}}$ through the
channel $U_{\mathcal{N}_{1}}^{A_{1}^{\prime}\rightarrow B_{1}E_{1}}$ gives
rise to a state $\phi^{A_{1}B_{1}E_{1}}$ where%
\[
\phi^{A_{1}B_{1}E_{1}}\equiv U_{\mathcal{N}_{1}}^{A_{1}^{\prime}\rightarrow
B_{1}E_{1}}(\phi^{A_{1}A_{1}^{\prime}}).
\]
The $n^{\text{th}}$ extensions of the above states, channel, and isometric
extension are respectively as follows:\ $\phi^{A_{1}^{n}A_{1}^{\prime n}}$,
$\phi^{A_{1}^{n}B_{1}^{n}E_{1}^{n}}$, $\mathcal{N}_{1}^{A_{1}^{\prime
n}\rightarrow B_{1}^{n}}$, and $U_{\mathcal{N}_{1}}^{A_{1}^{\prime
n}\rightarrow B_{1}^{n}E_{1}^{n}}$.

Alice also has access to her share $A_{2}^{n}$ of the state $\rho^{A_{2}%
^{n}B_{2}^{n}}$. There is another, more useful way of thinking about this
shared state. Let us first consider a purification $\psi^{A_{2}B_{2}E_{2}}$
of the state $\rho^{A_{2}B_{2}}$. We can think of the purification
$\psi^{A_{2}B_{2}E_{2}}$ as arising from sending a state $\psi^{A_{2}%
A_{2}^{\prime}}$ through a channel $\mathcal{N}_{2}^{A_{2}^{\prime}\rightarrow
B_{2}}$ with isometric extension $U_{\mathcal{N}_{2}}^{A_{2}^{\prime
}\rightarrow B_{2}}$:%
\[
\psi^{A_{2}B_{2}E_{2}}=U_{\mathcal{N}_{2}}^{A_{2}^{\prime}\rightarrow B_{2}%
}(\psi^{A_{2}A_{2}^{\prime}}).
\]
The tensor power state $\psi^{A_{2}^{n}B_{2}^{n}E_{2}^{n}}$ arises from
sending $n$ copies of the state $\psi^{A_{2}A_{2}^{\prime}}$ through the
tensor power channel $U_{\mathcal{N}_{2}}^{A_{2}^{\prime n}\rightarrow
B_{2}^{n}E_{2}^{n}}$. So, it is physically equivalent to say that Alice has
access to the system $A_{2}^{n}$ of the state $\psi^{A_{2}^{n}A_{2}^{\prime
n}}$ before the $A_{2}^{\prime n}$ system is transmitted through the channel
$U_{\mathcal{N}_{2}}^{A_{2}^{\prime n}\rightarrow B_{2}^{n}E_{2}^{n}}$, but
she does not have access to system $A_{2}^{\prime n}$.%

\begin{figure}
[ptb]
\begin{center}
\includegraphics[
natheight=3.953100in,
natwidth=7.666600in,
height=2.1819in,
width=2.9066in
]%
{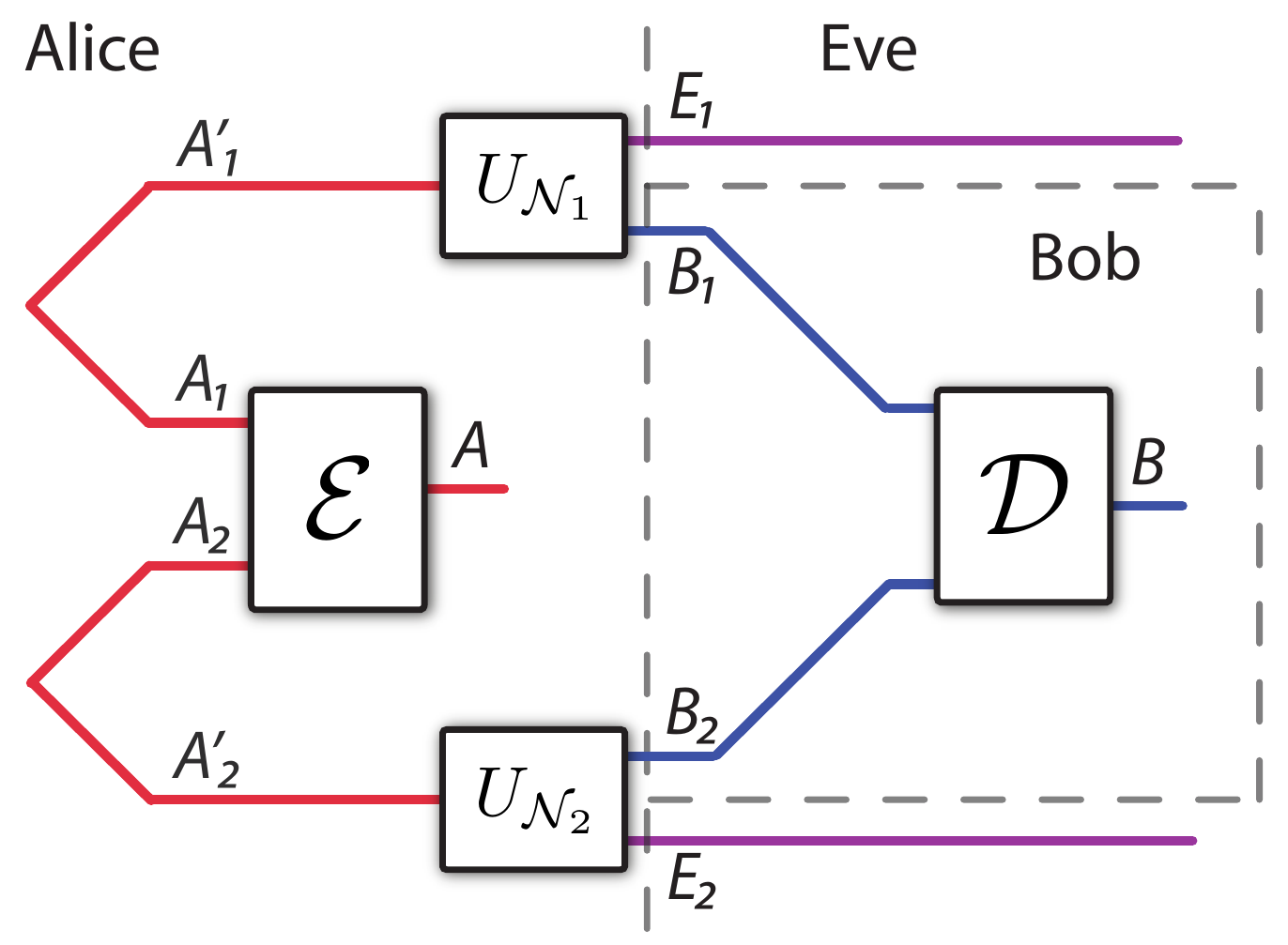}%
\caption{(Color online)\ A coding scheme for the channel-state protocol.}%
\label{fig:noisy-father}%
\end{center}
\end{figure}
\begin{figure}
[ptb]
\begin{center}
\includegraphics[
natheight=3.226600in,
natwidth=8.346300in,
height=1.299in,
width=3.1531in
]%
{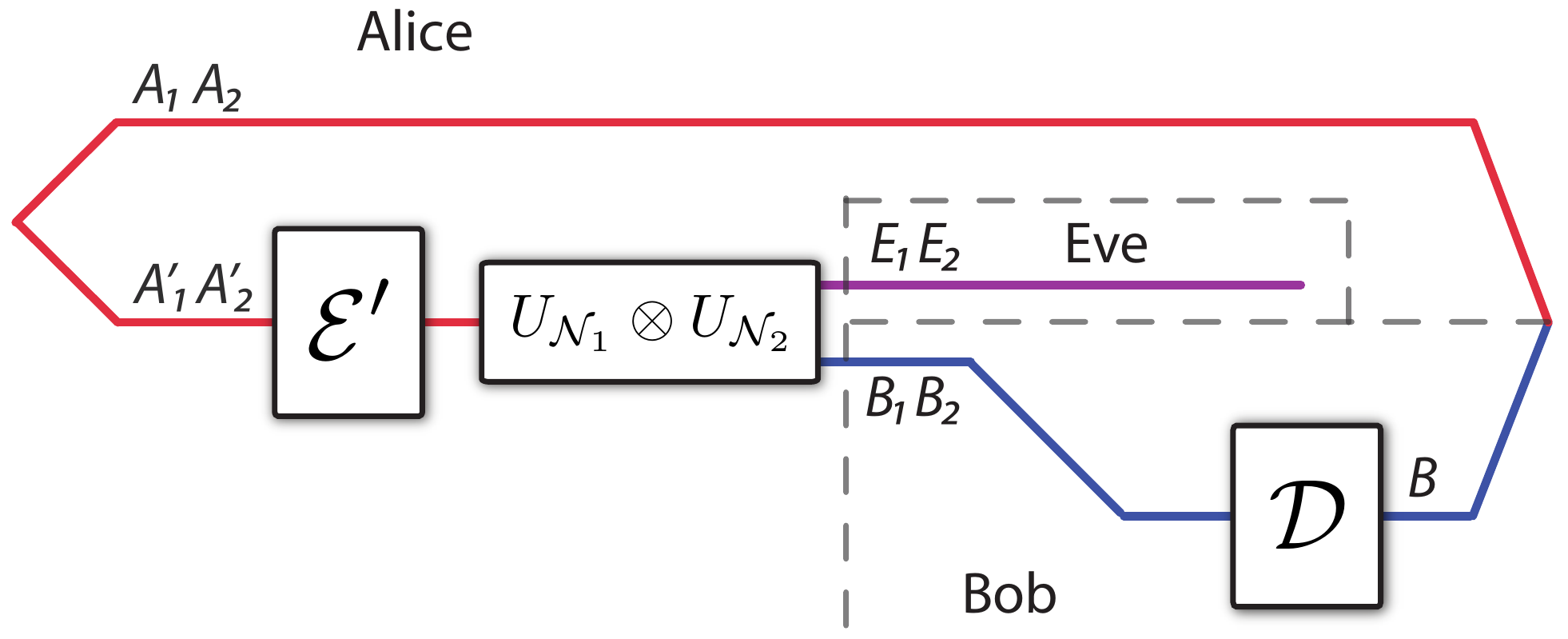}%
\caption{(Color online)\ Reduction of the coding scenario in
Figure~\ref{fig:noisy-father} to entanglement generation for the product
channel $U_{\mathcal{N}_{1}}\otimes U_{\mathcal{N}_{2}}$.}%
\label{fig:noisy-father-to-q-cap}%
\end{center}
\end{figure}

Alice prepares the state $\phi^{A_{1}^{n}A_{1}^{\prime n}}$ alongside the
state $\psi^{A_{2}^{n}A_{2}^{\prime n}}$. She performs an initial entangling,
isometric encoder $\mathcal{E}^{A_{1}^{\prime}A_{2}\rightarrow A_{1}^{\prime
}A_{2}}$ on each copy $\phi^{A_{1}A_{1}^{\prime}}\otimes\psi^{A_{2}%
A_{2}^{\prime}}$ of the state, so that the overall encoding is a tensor power
that we denote by%
\[
\mathcal{E}^{A_{1}^{\prime n}A_{2}^{n}\rightarrow A_{1}^{\prime n}A_{2}^{n}%
}(\phi^{A_{1}^{n}A_{1}^{\prime n}}\otimes\psi^{A_{2}^{n}A_{2}^{\prime n}}).
\]
She performs a \textit{typical subspace} measurement of the systems $A_{1}%
^{n}A_{2}^{n}$ followed by a \textit{type} measurement of the systems
\cite{qcap2008first,ADHW06FQSW}, ensuring that the systems $A_{1}^{n}A_{2}%
^{n}$ and $A_{1}^{\prime n}A_{2}^{\prime n}$ are maximally entangled. She
performs a random unitary $U$, selected from the Haar measure, on the systems
$A_{1}^{n}A_{2}^{n}$. This unitary is equivalent to applying the unitary
$U^{T}$ to the systems $A_{1}^{\prime n}A_{2}^{\prime n}$ because the systems
$A_{1}^{n}A_{2}^{n}$ and $A_{1}^{\prime n}A_{2}^{\prime n}$ are maximally
entangled \cite{ADHW06FQSW}. She then performs a projective measurement of the
systems $A_{1}^{n}A_{2}^{n}$ onto a subspace $S\subseteq A_{1}^{n}A_{2}^{n}$
of size $\left\vert S\right\vert $. This last measurement projects the systems
$A_{1}^{\prime n}A_{2}^{\prime n}$ onto a subspace $S^{\prime}\subseteq
A_{1}^{\prime n}A_{2}^{\prime n}$, where the subspace $S^{\prime}$ is
isomorphic to the subspace $S$ and the state on $SS^{\prime}$ is maximally
entangled \cite{qcap2008first}. Let $\mathcal{E}$ denote these encoding
operations on systems $A_{1}^{n}A_{2}^{n}$, and let $\mathcal{E}^{\prime}$
denote the equivalent operations occurring on systems $A_{1}^{\prime n}%
A_{2}^{\prime n}$. Figure~\ref{fig:noisy-father}\ gives a picture of this
protocol, and Figure~\ref{fig:noisy-father-to-q-cap}\ gives a picture of a
protocol that is formally equivalent to the one in
Figure~\ref{fig:noisy-father}. She sends Bob the result of the type
measurement (requiring only a sublinear amount of classical communication),
and she sends the result of the second projective measurement, requiring
$nI\left(  A_{1}A_{2};E_{1}E_{2}\right)  $ bits of classical communication
\cite{DW03c,DW03b}, so that he knows in which subspace the entanglement is encoded.

We now can exploit the simplified proof of the quantum coding theorem
(Theorem~1 of Ref.~\cite{qcap2008first}). Bob can perform a reliable decoding (resulting from a
decoupling of Bob's outputs $B_{1}^{n}B_{2}^{n}$ from Eve's outputs $E_{1}%
^{n}E_{2}^{n}$) if the size $\left\vert S\right\vert $ of the subspace $S$ is
not too large \cite{qcap2008first}: $\left\vert S\right\vert <2^{n\left(
H\left(  B_{1}B_{2}\right)  -H\left(  E_{1}E_{2}\right)  \right)  }$. The rate
of this code is%
\begin{align*}
\frac{\log\left\vert S\right\vert }{n} &  <H\left(  B_{1}B_{2}\right)
-H\left(  E_{1}E_{2}\right)  \\
&  =H\left(  B_{1}B_{2}\right)  -H\left(  A_{1}A_{2}B_{1}B_{2}\right)  \\
&  =I\left(  A_{1}A_{2}\rangle B_{1}B_{2}\right)  .
\end{align*}
We can maximize over all input states $\phi^{A_{1}A_{1}^{\prime}}$ and
isometric encoders $\mathcal{E}^{A_{1}^{\prime}A_{2}\rightarrow A_{1}^{\prime
}A_{2}}$ to have a code that achieves the one-shot capacity:%
\[
\max_{\phi,\mathcal{E}}I\left(  A_{1}A_{2}\rangle B_{1}B_{2}\right)  ,
\]
where the coherent information is with respect to the following state:%
\[
\mathcal{N}^{A_{1}^{\prime}\rightarrow B_{1}}(\mathcal{E}^{A_{1}^{\prime}%
A_{2}\rightarrow A_{1}^{\prime}A_{2}}(\phi^{A_{1}A_{1}^{\prime}}\otimes
\rho^{A_{2}B_{2}})).
\]
The maximization with respect to input states and encoders is equivalent to
performing a maximization over isometric preparations.

This code achieves the one-shot capacity in (\ref{eq:one-shot-cap}). We can
then block the channels and states together to give a superchannel
$\mathcal{N}^{A_{1}^{\prime l}\rightarrow B_{1}^{l}}$ and superstate
$\rho^{A_{2}^{l}B_{2}^{l}}$. Applying the above proof to this scenario gives a
code that achieves the capacity in Theorem~\ref{thm:NFP}.%

\section{Special Case}

\label{sec:special-case}We can restrict the above protocol to obtain a special
case. Suppose Alice prepares a state $\Phi^{A_{1}^{n}A_{1}^{\prime}}$ and
limits the preparation to be of the form $\mathcal{E}^{A_{1}^{\prime
}\rightarrow A^{\prime n}}\otimes\Lambda^{A_{2}^{n}}$, where the map
$\mathcal{E}^{A_{1}^{\prime}\rightarrow A^{\prime n}}$ is a quantum channel
encoding and the map $\Lambda^{A_{2}^{n}}$ is a quantum instrument
\cite{Yard05a}\ for entanglement distillation. Bob's decoding consists of an
operation $\mathcal{D}^{B_{1}^{n}\rightarrow B_{1}^{\prime}}\otimes
\Lambda^{B_{2}^{n}}$, where the map $\mathcal{D}^{B_{1}^{n}\rightarrow
B_{1}^{\prime}}$ is a quantum channel decoding and the map $\Lambda^{B_{2}%
^{n}}$ is a map that uses the classical information sent by Alice. This
protocol uses the forward classical channel for entanglement distillation
\cite{BDSW96}\ and uses the quantum channel for channel coding the system
$A_{1}^{\prime}$ only. The one-shot entanglement generation capacity for this
restricted scenario is%
\[
E^{(1)}(\mathcal{N}\otimes\rho)=\max_{\phi^{A_{1}A_{1}^{\prime}}}I\left(
A_{1}\rangle B_{1}\right)  +I\left(  A_{2}\rangle B_{2}\right)
\]
and is equal to the sum of the entanglement generation capacity with the
entanglement distillation capacity. This protocol is not optimal because it is
less than or equal to the one-shot capacity in (\ref{eq:upper-bound}).

\section{Resource Inequalities}

\label{sec:mother-father}We discuss some resource inequalities that follow
from the channel-state resource inequality in (\ref{eq:resource-inequality}).
We can generate a \textquotedblleft fully quantum\textquotedblright\ resource
inequality, by applying rule I from Ref.~\cite{DHW05RI}\ to the resource
inequality in (\ref{eq:resource-inequality}). We can apply rule I because the
communicated classical information is coherently decoupled. The resulting
resource inequality resembles the mother resource inequality:%
\[
\left\langle \mathcal{N}\otimes\rho\right\rangle +\frac{1}{2}I\left(
A_{1}A_{2};E_{1}E_{2}\right)  _{\omega}\left[  q\rightarrow q\right]
\geq\frac{1}{2}I\left(  A_{1}A_{2};B_{1}B_{2}\right)  _{\omega}\left[
qq\right]  .
\]

There is also a sense in which the mother and father protocol in
Refs.~\cite{DHW03,DHW05RI}\ arise as special cases of the channel-state coding protocol.
First, suppose that the state $\rho^{\otimes n}$ is equivalent to a rate $E$
maximally entangled state $\Phi^{A_{2}^{n}B_{2}^{n}}$ with $nE$ ebits of
entanglement where $E\geq I\left(  A_{1};E_{1}\right)  /2$. Then, it is
best to act with the father protocol. The resource inequality is equivalent
to that for the father protocol (modulo some classical communication):%
\[
\left\langle \mathcal{N}\right\rangle +\frac{1}{2}I\left(  A_{1};E_{1}\right)
\left[  qq\right]  \geq\frac{1}{2}I\left(  A_{1};B_{1}\right)  \left[
q\rightarrow q\right]  .
\]

Another special case of this protocol is the mother protocol. Suppose that the
channel $\mathcal{N}$ is a noiseless qubit channel id$^{A_{1}\rightarrow
B_{1}}$ of rate $Q$ where $Q\geq I\left(  A_{2};E_{2}\right)  /2$. Then the
resource inequality is equivalent to that for the mother protocol:%
\[
\left\langle \rho\right\rangle +\frac{1}{2}I\left(  A_{2};E_{2}\right)
\left[  q\rightarrow q\right]  \geq\frac{1}{2}I\left(  A_{2};B_{2}\right)
\left[  qq\right]  .
\]

\section{Superactivation}

We now discuss how the channel-state coding protocol can lead to a
superactivation effect. The main finding in Re\.{f}.~\cite{science2008smith}
was the following inequality:%
\[
\frac{1}{2}P\left(  \mathcal{N}\right)  \leq Q\left(  \mathcal{N}%
\otimes\mathcal{A}\right)  ,
\]
where $P\left(  \mathcal{N}\right)  $ denotes the private capacity\ of a
channel $\mathcal{N}$ \cite{Devetak03,CWY04} and $Q\left(  \mathcal{N}%
\otimes\mathcal{A}\right)  $ denotes the joint quantum capacity of the channel
$\mathcal{N}$ and a symmetric channel $\mathcal{A}$ (a symmetric channel is
one that behaves the same under interchange of its receiver and its
environment, and thus has zero quantum capacity by a no-cloning argument
\cite{SSW08}). Smith and Yard showed that an entanglement-binding channel
$\mathcal{N}$ \cite{H3O05}\ , one that has zero quantum capacity but non-zero
private capacity, and a 50\% erasure channel \cite{Grassl:1997:33}, an example
of a symmetric channel, can combine to have a non-zero quantum capacity.

Devetak showed that the secret key generation capacity $K\left(
\mathcal{N}\right)  $ of quantum channel $\mathcal{N}$\ is equal to its
privacy capacity $P\left(  \mathcal{N}\right)  $, and he also showed that its
entanglement generation capacity $E\left(  \mathcal{N}\right)  $ is equal to
its quantum capacity $Q\left(  \mathcal{N}\right)  $ \cite{Devetak03}. Thus,
it is possible to translate the above Smith-Yard inequality as follows:%
\[
\frac{1}{2}K\left(  \mathcal{N}\right)  \leq E\left(  \mathcal{N}%
\otimes\mathcal{A}\right)  .
\]
The question now is whether we can have the following inequality for a noisy
bipartite state $\rho^{AB}$ and a noisy erasure channel $\mathcal{A}$:%
\begin{equation}
\frac{1}{2}K\left(  \rho\right)  \leq E\left(  \rho\otimes\mathcal{A}\right)
. \label{eq:superactivation}%
\end{equation}
An example of the above inequality for our scenario follows directly from the
example of Smith and Yard in the appendix of Ref.~\cite{science2008smith}.
Suppose that we have the state $\left\vert \rho\right\rangle ^{XAC}$ on page 3
of the Supplementary Materials in Ref.~\cite{science2008smith}. Let us relabel
this state as $\left\vert \rho\right\rangle ^{XA_{2}C}$. Sending the $A_{2}$
system through an entanglement-binding channel gives a state $\left\vert
\rho\right\rangle ^{XB_{2}E_{2}C}$. Discarding the $C$ system leads to a state
$\rho$ on $XB_{2}$, where Alice possesses $X$ and Bob possesses $B_{2}$. Alice
and Bob can distill some secret key from this state ($K\left(  \rho\right)
>0$), but cannot distill any maximal entanglement ($E\left(  \rho\right)
=0$). Suppose now that the state is $\left\vert \rho\right\rangle
^{XB_{2}E_{2}C}$. Alice can send the $C$ system through a 50\% erasure
channel. By the same proof method in Ref.~\cite{science2008smith}, it is
possible to show the inequality in (\ref{eq:superactivation}) for this
particular setup. The key to this modification of the Smith-Yard proof is that
there is entanglement between Alice's systems $X$ and $C$, implying that an
entangling encoder in the channel-state coding protocol outperforms a strategy
such as the one in Section~\ref{sec:special-case}, which is not able to
generate any entanglement for this state. Thus, we have a superactivation
effect occurring for the channel-state coding protocol.

\section{Conclusion}

We have introduced a new protocol, the channel-state coding protocol, that
combines a noisy quantum channel with a noisy quantum state to generate
entanglement. This protocol
performs well when entanglement is not perfect and should aid
in the effort to examine entanglement-assisted codes
with imperfect entanglement. The channel-state coding protocol also
exhibits the superactivation effect, where a state with no distillable
entanglement and a zero-capacity quantum channel can generate
maximal entanglement. An open task is to construct a protocol that achieves quantum communication rather
than mere entanglement generation.

M.M.W. thanks Andreas Winter for useful discussions and
acknowledges research 
grant SAIC-1669,  the National Research Foundation \&
Ministry of Education, Singapore, and the Centre for Quantum Technologies.

\bibliographystyle{IEEEtran}
\bibliography{Ref}

\end{document}